\documentclass[11pt]{article}
\usepackage{moriond,epsfig}

\def\Journal#1#2#3#4{{#1} {\bf #2}, #3 (#4)}

\def\PLB{{\em Phys. Lett.}  B}

\def\PRD{{\em Phys. Rev.} D}

\def\be{\begin{equation}}
\def\ee{\end{equation}}
\def\bea{\begin{eqnarray}}
\def\eea{\end{eqnarray}}

\begin{document}
\vspace*{4cm}
\title{STRUCTURE FUNCTIONS AND EXTRACTION OF PDFS AT HERA}

\author{ N. RAI\v CEVI\' C for the H1 and ZEUS collaborations}

\address{Department of Physics, Faculty of Science, University of Montenegro,\\
Cetinjski put BB, P.O. Box 211, 81000 Podgorica, Montenegro}

\maketitle\abstracts{
Results from the HERA experiments, H1 and ZEUS, on $e^\pm p$ deep inelastic 
scattering (DIS) provide an important contribution to the knowledge of the 
proton structure and QCD. The data were collected in the  years 1994-2000 
(HERA I) and 2003-2006 (HERA II) in the center-of-mass energy 
of $\sqrt{s} \;= \; 300$ GeV in 1994-1997 and 319 GeV from 1998. During the 
HERA II period, the lepton beams were longitudinally polarised. The most 
recent results on neutral current (NC) and charge current (CC) DIS cross 
sections from HERA II data are presented. Results on Parton Density Functions 
(PDFs) and 
the strong coupling $\alpha _s$ extracted from HERA I data are discussed. 
Accounting for the correlation of electroweak parameters with PDFs, a combined 
electroweak and QCD analysis is performed for the first time at HERA.} 

\section{Structure Functions and Cross Sections }

\begin{figure}
\rule{0.7cm}{0.0mm}
a)
\psfig{figure=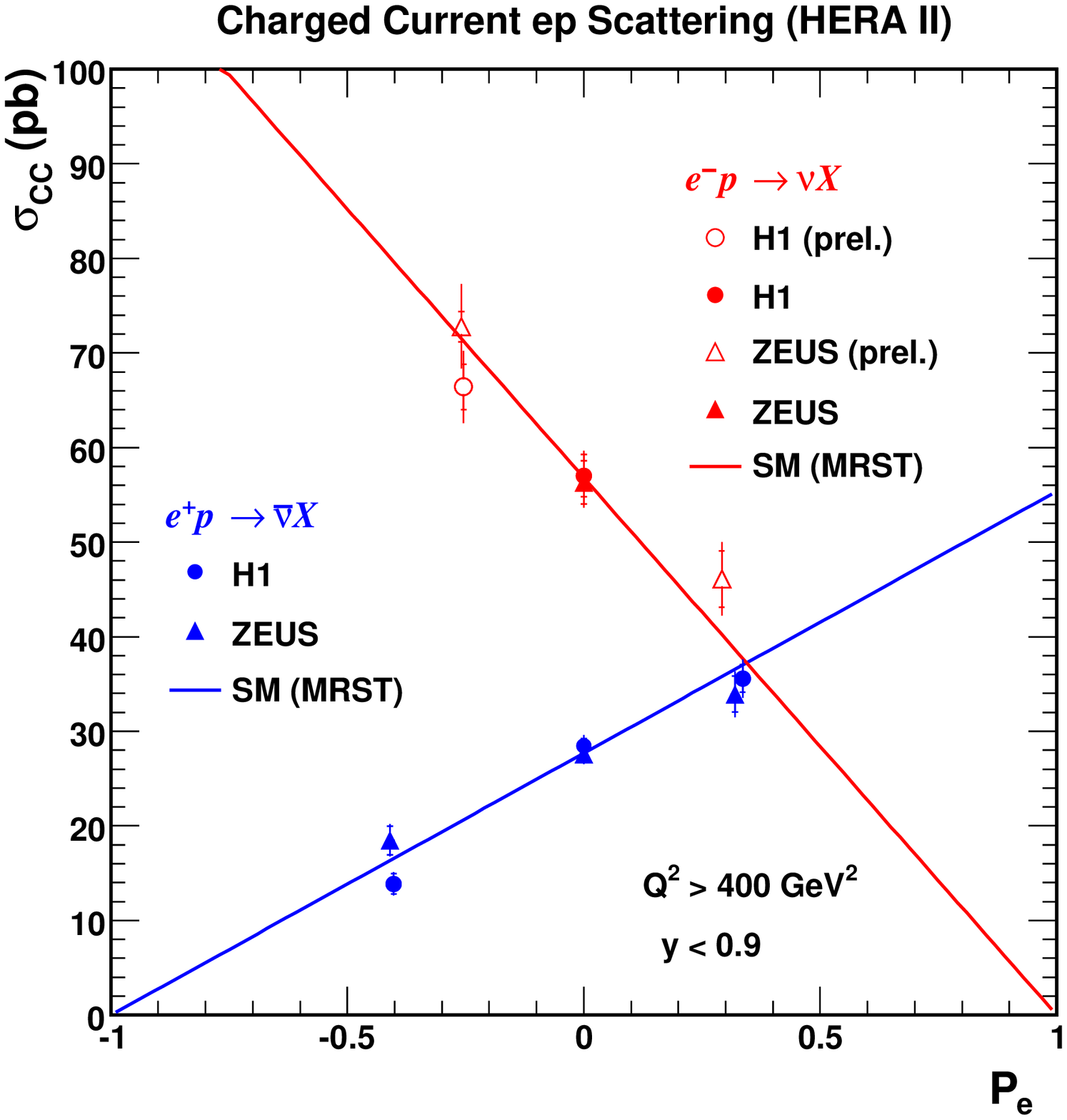,height=2.5in}
\rule{0.5cm}{0.0mm}
b)
\psfig{figure=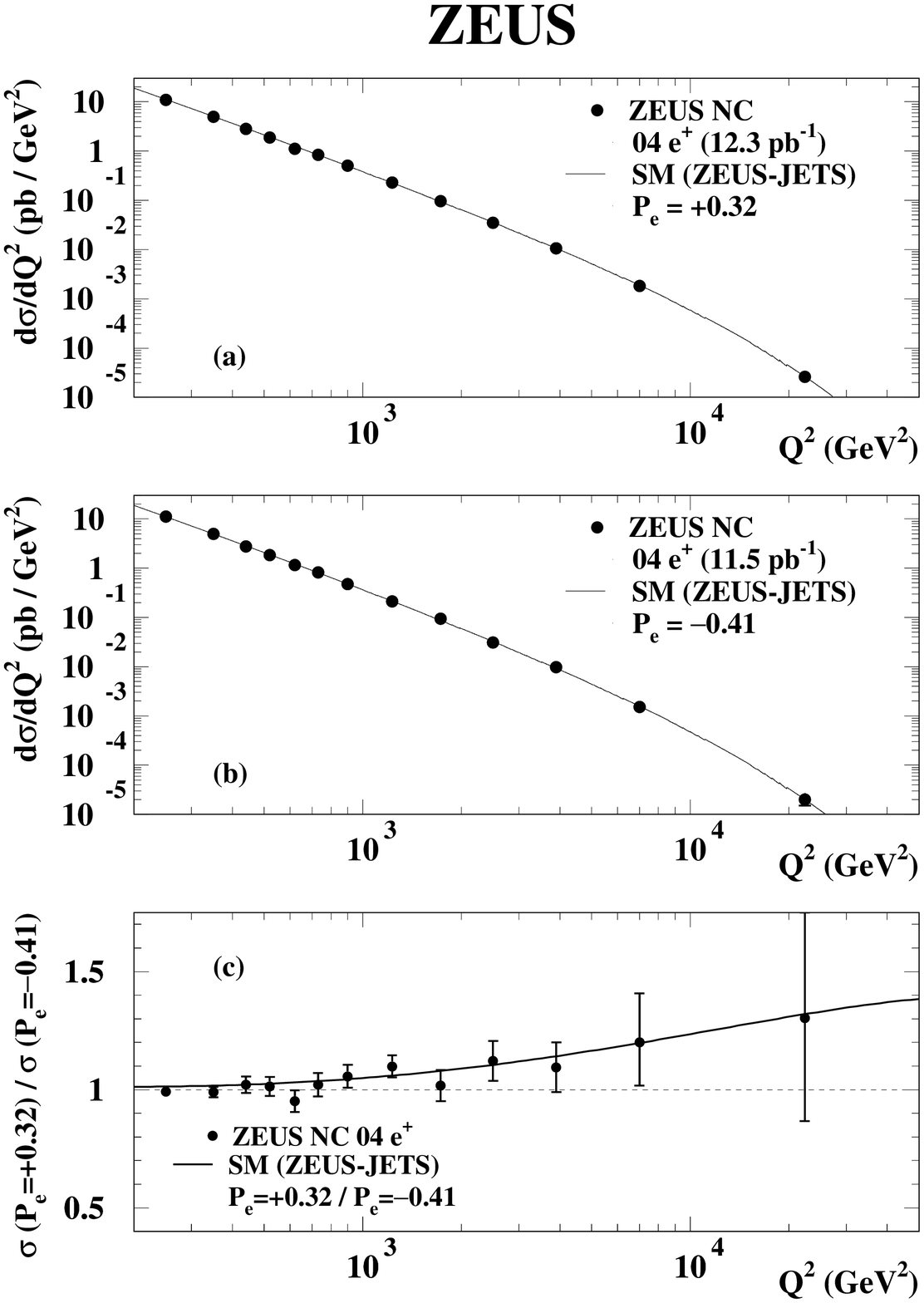,height=3.5in}
\caption{a) $e^ \pm p$ CC cross sections measurements from HERA for
different lepton polarisations. b) $e^+ p$ NC differential cross sections, 
$d \sigma / dQ^2$, obtained from ZEUS HERA II data.
\label{fig:figure1}}
\end{figure}

NC DIS processes proceed via exchange of photons 
and $Z^0$ bosons. Photon exchange dominates and its contribution to
the cross section is given in terms of the proton structure function
F$_2(x,Q^2)$ which provides information on the total quark content of the
proton at given values of the Bjorken scale variable $x$ and of the modulus
of the squared four-momentum transfer, $Q^2$, carried by the exchanged boson. 
Both HERA experiments have confirmed $Q^2$ evolution of $F_2$ predicted by 
perturbative QCD (pQCD) over five orders of magnitude in  $Q^2$ and 
x~\cite{v1,v2,v3,v4,v5}. At low
and medium $Q^2$ precision of $F_2$ is 2-3 $\%$ while at high $Q^2$ region
is statistically limited.\\
At $Q^2 \ge {M_Z}^2$, NC cross section for $e^+p$ and $e^-p$ scattering 
differ due to the electroweak effects, especially the $\gamma-Z$
interference. The difference is described by the proton structure function 
$xF_3(x,Q^2)$ which in pQCD is given by the difference between the quark and
anti-quark density functions, thus providing information on the valence quark 
contribution in the proton. Uncertainties of the existing $xF_3$ measurements
are dominated by the limited statistics of the $e^-p$ sample.\\ 
\noindent Longitudinal structure function, $F_L$, is identically zero in
lowest order QCD, but due to gluon radiation gets a non-zero value in
pQCD. Measurements of $F_L$ can thus provide constraints on the gluon PDF
which are complementary to that obtained from the scaling violations of $F_2$
assuming DGLAP evolution~\cite{v11}. 
Indirect measurements of $F_L$ performed by H1 collaboration suggest that
$F_L$ remains non-zero down
to the lowest $Q^2$ values measured~\cite{v1,v2,v6}. Significant progress in
$F_L$ measurements at HERA can only be made by reducing the proton beam energy
which provides its direct measurements~\cite{v7}.

CC DIS processes are mediated by $W^\pm$ bosons and  provide
complementary information about partonic composition of proton since they are
sensitive to particular quark flavor with the certain charges to couple to
the exchanged boson.

To increase luminosity, an upgrade 
of HERA in the H1 and ZEUS detector regions was performed in 2001. In order to
increase the $e^- p$ statistics, HERA has been running with $e^ -p$ beams
since December 2004 and is expected to finish in summer 2006, when 
$e ^ + p$ mode should start. Luminosity of HERA II $e^- p$ sample is about 150
$pb^ -1$ roughly per experiment which is already about factor of 10 greater
then achieved during HERA I period. End of HERA operation is planned for 
summer 2007.\\
\noindent First results for the cross sections for charged
and neutral current deep inelastic scattering in $e^\pm p$ collisions with a 
longitudinally polarised lepton beams~\cite{v8,v9} have been obtained 
at HERA.
The CC cross section measurements depending on polarisation are
presented on figure~\ref{fig:figure1}a. In the Standard Model (SM) only 
left-handed electrons 
and right-handed positrons take part in CC interactions and the CC cross 
section depends linearly on the polarisation P as: 
$\sigma ^{CC} (P) \; = \; (1+P) \sigma ^{CC} (0)$.
As can be seen from the figure, the data are found to be consistent with the
absence of right handed charged currents as predicted by the SM. 
\noindent NC interactions are also sensitive to the lepton polarisation.
Electromagnetic contribution which dominates at low $Q^2$ does not depend
on polarisation. Polarisation dependence ocurres mainly via interference 
between $\gamma$ and Z boson exchanges. Figure~\ref{fig:figure1}b shows 
ZEUS measurements of 
the differential cross section $d \sigma / d Q^2$ for the NC DIS for positive 
and negative longitudinal polarisations and the ratio of the two cross
sections. The measurements are consistent with the SM predictions evaluated 
using the ZEUS-JETS PDFs
(described in the next section) and are also consistent with the expectations
of the electroweak SM for polarised NC DIS.
  
\section{Determination of Parton Densities and $\alpha _s$ }

\noindent The H1~\cite{v1,v3} and ZEUS~\cite{v4,v10} collaborations have 
performed QCD fits to extract parton densities using various combination
of HERA and other data. The fits are based on the evolution of the PDFs
with $Q^2$ using DGLAP equations in Next-to-Leading-Order
(NLO).
\begin{figure}
\rule{1cm}{0.0mm}
a)
\psfig{figure=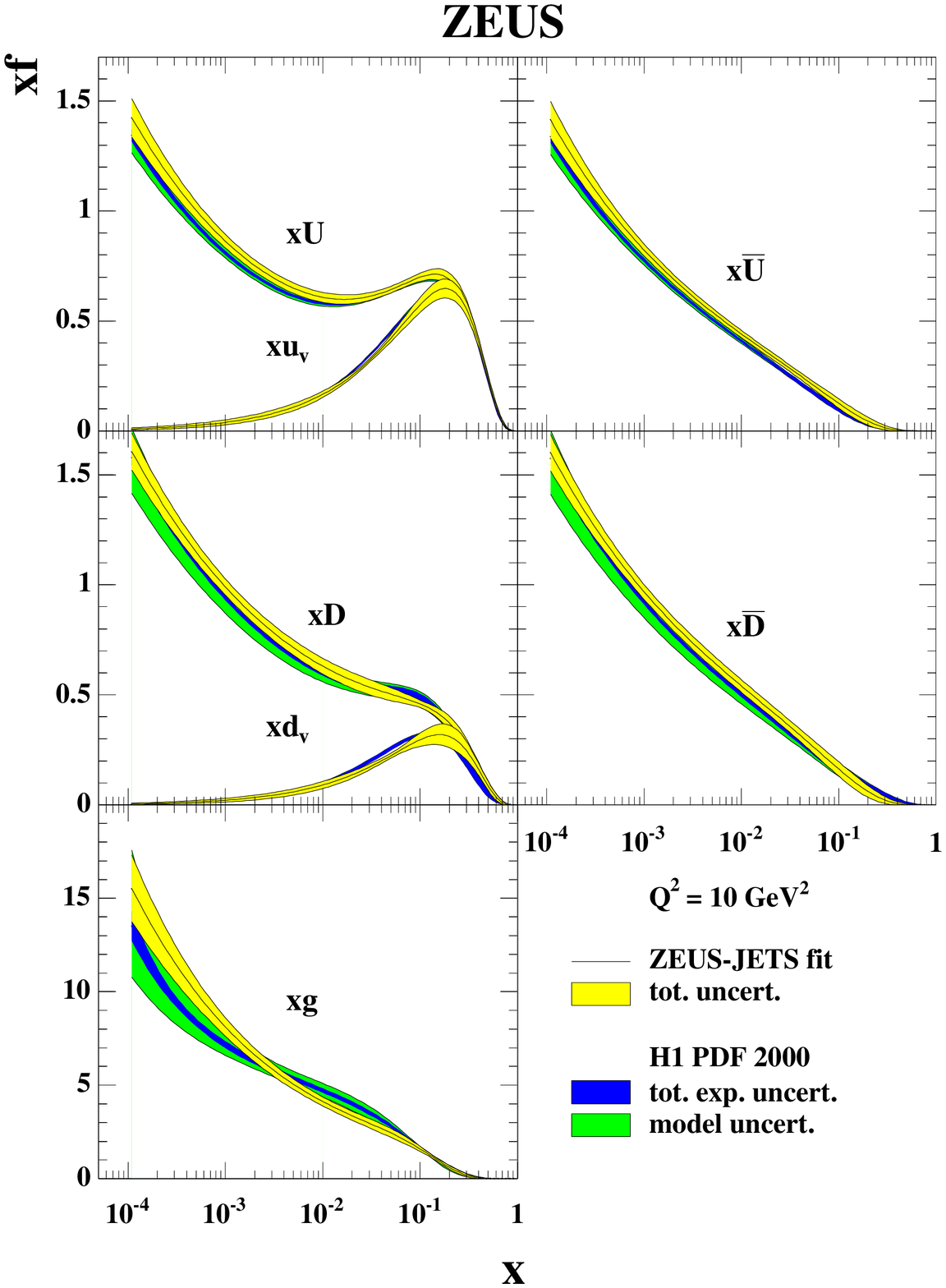,height=3.50in}
\rule{0cm}{0.0mm}
\vskip -8.70cm
\hskip 9.0cm
b)
\epsfig{figure=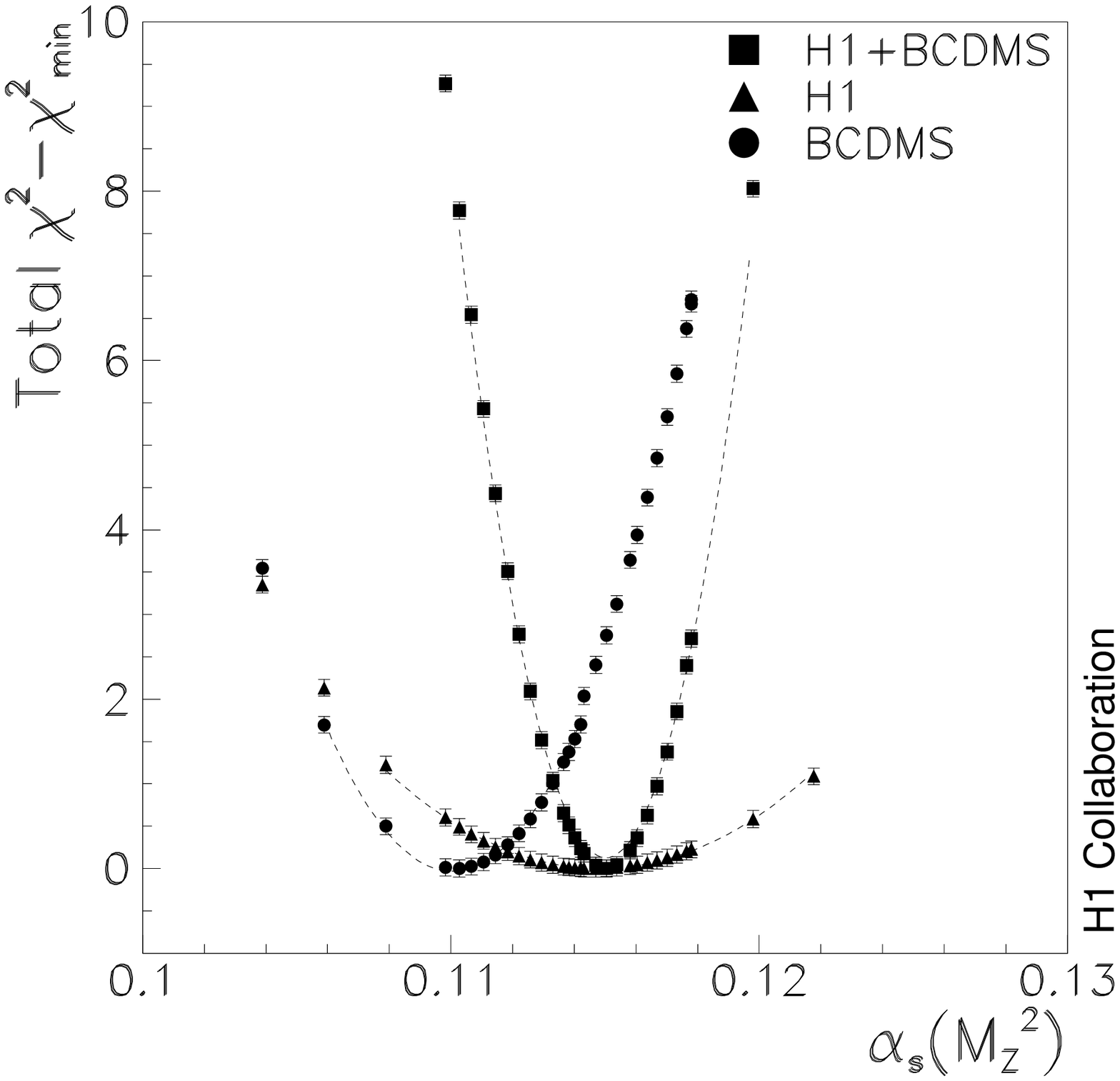,height=1.5in}
\rule{-1cm}{0.0mm}
\vskip -0.1cm
\hskip 8.7cm
c)
\psfig{figure=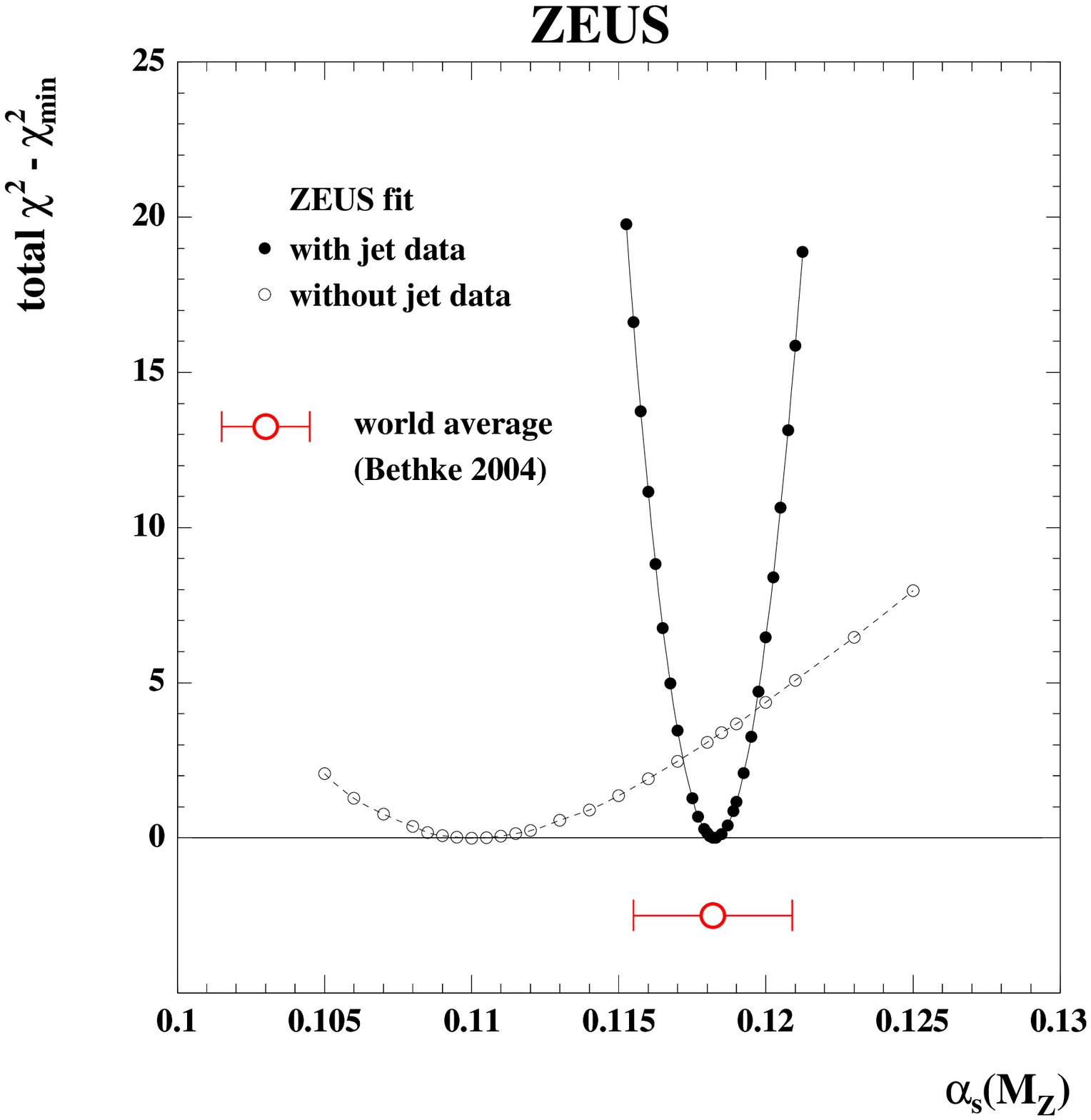,height=1.7in}
\caption{a) PDFs obtained from H1PDF2000 and ZEUS-JETS fits. b) $\chi ^2$
for $\alpha _s$ fits to the H1 $ep$ and BCDMS $\mu p$ data separately and for
the fit using data of the two experiments combined.
c) $\chi ^ 2$ profiles of the $\alpha _s$ ZEUS fits with and without jet data.
\label{fig:figure2}}
\end{figure}  
\noindent The precision and kinematic coverage of the H1 and ZEUS data allow
to perform QCD analysis (H1PDF2000~\cite {v3}, ZEUS-JETS~\cite{v10} fits) 
with HERA data alone. A fit within one $ep$ experiment provides better control
of the systematic uncertainties and avoids many theoretical uncertainties, 
arising from heavy target correction, higher-twist
contribution or isospin symmetry assumptions. Currently fits based on HERA
data only are limited by statistics for high-$x$ domain and significant
improvements are expected with inclusion of HERA II data. \\
\noindent In ZEUS-JETS fit, data on jet production were used to constrain the
mid to high-x gluon density. The four combinations of up and down PDFs for 
$U = u+c$, $\overline U = \overline u + \overline c$, $D = d + s$, 
$\overline D = \overline d + \overline s$ 
and the gluon density function are shown in figure~\ref{fig:figure2}a as 
functions of $x$ for fixed 
$Q^2 = 10$ $GeV^2$.  Valence densities are obtained from the previous
combinations of up-type and down-type  and their anti-quark-type distributions
as $u_v = U - \overline U$ and $d_v = D - \overline D$.   Results obtained by 
H1 and ZEUS are in fair agreement with each other. The residual differences
may originate from the different functional forms of the parameterisations of 
PDFs, constrains imposed on the
densities, phase space, $Q^2$ start scale, the treatment of heavy quarks, 
treatment of experimental uncertainties, the data sets used, etc.\\
\noindent The pQCD fits provide a precise determination of the running strong
coupling constant, $\alpha _s$, with the experimental uncertainty 2-3 $\%$. 
Determination of $\alpha _s$ in the H1 collaboration~\cite{v1} is obtained by 
combining the low-x data of H1 with $\mu p$ scattering data of the BCDMS collaboration
at large $x$ which reduces uncertainties of the measurements which can be seen
from figure~\ref{fig:figure2}b. The
additional constrain on the gluon PDF from the jet data in the ZEUS-JETS fit, 
has provided an improved determination of $\alpha _s$ which can be seen form
figure~\ref{fig:figure2}c. The uncertainty in $\alpha _s$ due to terms beyond 
NLO has
been estimated as $\Delta \alpha _s \simeq \pm 0.005$ by variation of
the choice of scales. 

\section{Combined Electroweak and QCD Fit}

\begin{figure}
\rule{2cm}{0.0mm}
\psfig{figure=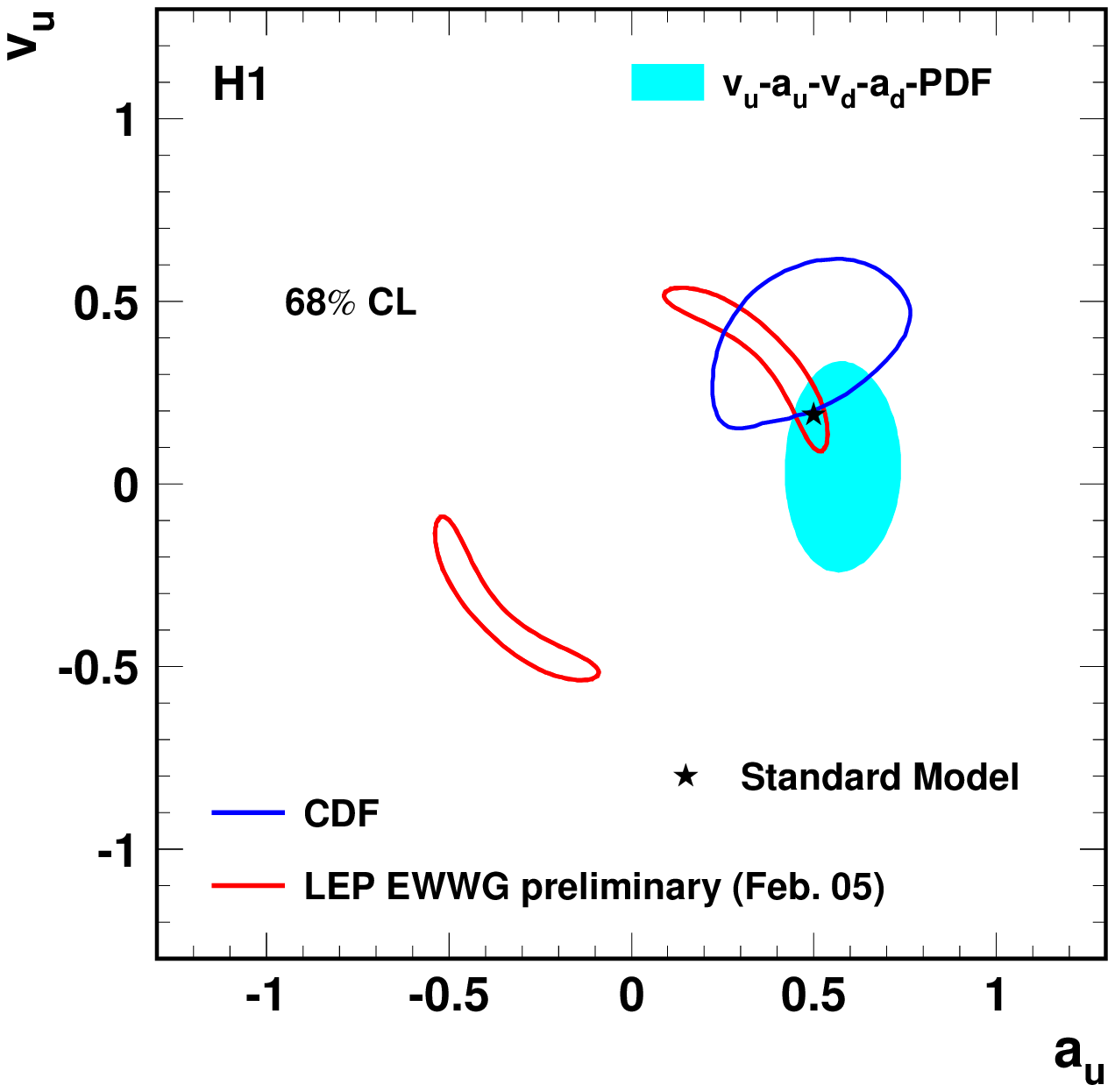,height=1.6in}
\rule{1.5cm}{0.0mm}
\psfig{figure=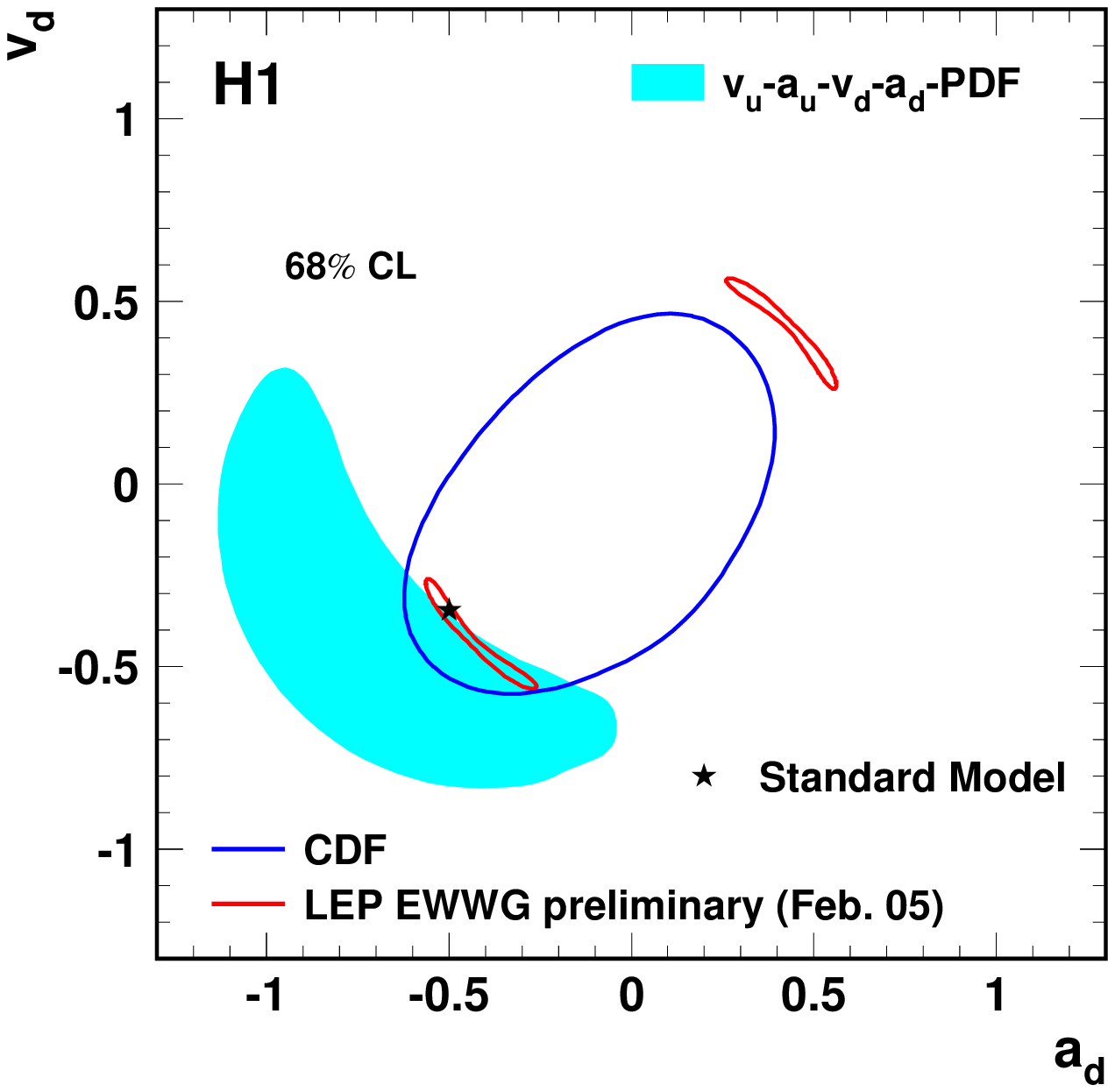,height=1.6in}
\caption{Results on the weak neutral current couplings of $u$ and 
$d$ quarks to the $Z^0$ boson from H1 compared with results from
the CDF, LEP and the SM expectation.
\label{fig:figure3}}
\end{figure}

The combined electroweak (EW) and QCD analysis~\cite{v12}, performed on data collected
by H1 during HERA I with luminosity of 117 $pb ^ {-1}$, follow the same
procedure as used in H1PDF2000 fit~\cite{v3}.
Taking into account dependence of the CC cross section on $Q^2$, the
propagator mass has been measured within the SM framework. The $W$ mass has 
been also measured in the on-mass-shell scheme. 
The NC data have been used to extract the weak vector and axial-vector
couplings of $u$ and $d$ quarks ($v_u$, $a_u$, $v_d$, $a_d$) to the
$Z^0$ boson  . In the
fit $v_u - a_u - v_d - a_d - PDFs$, the vector and axial-vector couplings are
treated as free parameters. The results at 68 $\%$ confidence level (CF) are 
shown in figure~\ref{fig:figure3} together with
results from CDF~\cite{v13} and preliminary results from LEP~\cite{v14}.
Results expected from the SM are also shown.  
Precision of HERA measurements is comparable to that from the
CDF. These measurements are sensitive to $u$ and $d$ quarks separately and
also resolve sign ambiguity of LEP measurement.  

\end{document}